\documentclass[11pt]{article}
\usepackage{graphicx}
\usepackage{amsmath}
\usepackage{amsfonts}
\usepackage{amssymb}
\usepackage{epsfig}
\usepackage{times}
\usepackage{cite}
\usepackage{calc}
\usepackage{version}
\usepackage[english]{babel}
\usepackage{epsf}
\usepackage{graphics}
\usepackage{caption}
\usepackage{upgreek}
\usepackage{ulem}
\usepackage[dvips]{color}

\begin{document}

\begin{titlepage}

\null

\vskip 1.5cm

\vskip 1.cm

{\bf\large\baselineskip 20pt
\begin{center}
\begin{Large}
Inclusive transverse momentum distribution of hadrons in jets produced in PbPb and pp collisions at the LHC: Data
versus jet-quenching Monte Carlos
\end{Large}
\end{center}
}
\vskip 1cm

\begin{center}
Redamy P\'erez-Ramos\footnote{
Sorbonne Universit\'e, UPMC Univ Paris 06, UMR 7589, LPTHE, F-75005, Paris, France}
\footnote{
CNRS, UMR 7589, LPTHE, F-75005, Paris, France}
\footnote{
Postal address: LPTHE tour 13-14, $4^{\text{\`eme}}$ \'etage, UPMC Univ Paris 06, BP 126, 4 place Jussieu, F-75252 Paris Cedex 05 (France)}
\footnote{
Department of Physics, P.O. Box 35, FI-40014 University of Jyv\"askyl\"a, Jyv\"askyl\"a, Finland}
\footnote{
e-mail: redamy.r.perez-ramos@jyu.fi, perez@lpthe.jussieu.fr}
\end{center}

\baselineskip=15pt

\vskip 3.5cm

The inclusive transverse momentum ($p_t$) distribution of hadrons inside jets 
produced in PbPb and pp collisions are simulated with the {\sc YaJEM} 
and {\sc pythia6} Monte Carlo (MC) event generators. The effects of jet quenching
are studied via the ratios of PbPb over pp hadron $p_t$ spectra, either by 
accounting for the induced virtuality $\Delta Q^2$ transferred from the strongly-interacting medium 
to the parton shower or by modifying the soft sector of the parton-to-hadron fragmentation functions. 
The MC results are compared to experimental jet data measured by the CMS experiment at a center-of-mass energy of 2.76 TeV 
in four jet $p_T^{\rm jet}$ ranges above $100$ GeV, accounting or not for the experimental jet
reconstruction biases. The level of data-MC (dis)agreement provides valuable
information on the mechanism of parton energy loss.

\end{titlepage}

\section{Introduction}

High-energy heavy-ion collisions provide the means to study the properties of QCD matter in the quark-gluon
plasma (QGP) state.
Highly virtual partons produced in such collisions experience a strong energy degradation as they travel 
through the strongly-interacting medium, resulting in the suppression of high transverse momentum 
leading hadrons \cite{Adcox:2001jp,Adams:2003kv,Aamodt:2010jd,CMS:2012aa} and 
jets~\cite{Chatrchyan:2013kwa,Chatrchyan:2014ava}. The study of such ``jet quenching'' phenomena
provides information on the thermodynamical and transport properties of the QGP~\cite{d'Enterria:2009am}.
Since the start of the heavy-ion program at the LHC, with its plethora of new hard observables
available, new theoretical approaches have been developed to study the 
interaction of energetic partons with the hot and dense medium~\cite{Abreu:2007kv}.
In particular, realistic Monte Carlo (MC) codes for the simulation of in-medium parton shower evolution have
been constructed from ``QCD vacuum'' event 
generators~\cite{Zapp:2008gi,Lokhtin:2005px,Renk:2008pp,Renk:2009nz,Armesto:2008zza,Armesto:2009fj} 
such as {\sc pythia}~\cite{Bengtsson:1986hr,Norrbin:2000uu} 
and {\sc herwig}~\cite{Corcella:2000bw}. {\sc jewel} 
(Jet Evolution With Energy Loss)~\cite{Zapp:2008gi}
implements elastic and inelastic medium interactions which lead to distinctive 
modifications of the jet fragmentation pattern, including Landau-Pomeranchuk-Migdal 
destructive interference effects in a probabilistic framework~\cite{Baier:1996sk}. In 
{\sc q-pythia}~\cite{Armesto:2008zza,Armesto:2009fj}, medium effects are introduced via 
an extra term in the QCD splitting functions arising from the multiple-soft 
scattering approximation. In {\sc pyquen}~\cite{Lokhtin:2005px}, gluon 
radiation is associated with each parton scattering in 
the hot and dense medium and interference effects are included through the 
modified radiation spectrum as a function of the medium temperature (i.e. see~\cite{Lokhtin:2014vda}
for the computation of fragmentation functions with the {\sc pyquen} MC). 
The MC used in this work, {\sc YaJEM}
(Yet Another Jet Energy-loss Model)~\cite{Renk:2008pp,Renk:2009nz}, assumes in its default setup that 
the virtuality of partons interacting with the medium increases according
to the medium transport coefficient connected to the virtuality
gain per unit pathlength. Other extensions of the {\sc YaJEM} code such as
{\sc YaJEM-DE} and {\sc YaJEM-E}~\cite{Renk:2014lza} simulate a showering 
process evolved down to a hadronization scale which depends on the parton's energy and
medium pathlength.

In the present paper we compute the inclusive transverse momentum ($p_t$) distribution
of hadrons inside jets $dN/dp_t$ with the {\sc YaJEM} and 
{\sc pythia6}~\cite{Bengtsson:1986hr,Norrbin:2000uu} codes. We construct the ratios of PbPb over pp spectra
and compare them to experimental jet data measured by the CMS experiment at a center-of-mass energy of 2.76 TeV. 
In the PbPb case, it is assumed that the cascade of branching partons traverses a 
medium characterized by a local transport coefficient $\hat{q}$ 
such that at the end the virtuality of the leading parton is increased by a 
total $\Delta Q^2$ factor which widens the phase space and leads to the jet quenching. 
We consider also an alternative scenario based on the Borghini-Wiedemann (BW) 
model~\cite{Borghini:2005em}, where the singular part of the branching probabilities 
in the medium is increased by a multiplicative factor 
$1+f_{\rm med}$, such that $P_{a\to bc}=(1+f_{\rm med})/z+{\cal O}(1)$,
where $a\to bc$ describes the QCD parton branchings, i.e. $q(\bar q)\to q(\bar q)g$ and 
$g\to gg$ with $g\to q\bar q$ unchanged. In this case, the jet quenching is 
described by the extra amount of medium-induced soft gluons ($f_{\rm med}>0$) 
as compared to the vacuum ($f_{\rm med}=0$) which widens the transverse jet shape. 
In both cases, the final parton-to-hadron transition takes place in the vacuum, 
using the Lund model~\cite{Andersson:1983ia}, for hadronization scales below $Q_0\approx 1$ GeV. 
 
Aiming at performing a realistic comparison of {\sc YaJEM} with the CMS data requires
following the CMS data analysis as closely as possible. For this purpose, 
jets are first reconstructed from all particles by using the anti-$k_t$ 
algorithm \cite{Cacciari:2011ma,Cacciari:2005hq,Cacciari:2008gp} with a resolution parameter 
$R=0.3$. Secondly, charged particles with $p_{t}>1$~GeV are selected and reclustered
within the $p_T^{\rm jet}$ ranges $100\leq p_T^{\rm jet}({\rm GeV})\leq 120$, $120\leq p_T^{\rm jet}({\rm GeV})\leq 150$, 
$150\leq p_T^{\rm jet}({\rm GeV})\leq 300$ and $100\leq p_T^{\rm jet}({\rm GeV})\leq 300$ reported 
by the CMS collaboration~\cite{Chatrchyan:2014ava}. The 
condition $p_{t}>1$ GeV removes a very large underlying-event background 
but may bias the jet study. In order to illustrate the role of the bias 
caused by the jet-finding procedure and, particularly, that required by the
CMS trigger which takes jets above $p_T^{\rm jet}\geq 100$ GeV, we compare the data-driven ``biased'' ratios with
the ``unbiased'' ratios obtained by analyzing the jets at the MC-truth level.
Note that we use different notations for the hadron's transverse momentum $p_t$, the final
reconstructed $p_T^{\rm jet}$, which is potentially lower ($p_T^{\rm jet}<p_T^{\rm parton}$) than the initial
parton $p_T^{\rm parton}$ as a result of the main biases such as (i) reconstruction for the jet resolution $R=0.3$, 
(ii) charged particle selection, (iii) soft background removal $p_t>1$ GeV 
and (iv) the aforementioned $p_T^{\rm jet}$ cuts.

\section{Monte Carlo analysis for the medium-modified $\boldsymbol{p_t}$ distribution of 
hadrons via the PbPb/pp ratios}

In the {\sc YaJEM} code, the hard-scattered partons evolving into a jet shower are embedded in a 
hydrodynamical medium whose transport coefficient is taken to be~\cite{Renk:2011gj}, 
\begin{equation}
\label{eq:qhat}
\hat{q}(\zeta)=K\cdot2\cdot\epsilon^{3/4}(\zeta)F(\rho(\zeta),\alpha(\zeta))
\end{equation}
with
$$
F(\rho(\zeta),\alpha(\zeta))=\cosh\rho(\zeta)-\sinh\rho(\zeta)\cos\alpha(\zeta),
$$
where $\epsilon$ is the local energy density of the hydrodynamical medium, 
$F$ is a hydrodynamical flow correction factor accounting for the Lorentz contraction 
of the scattering centers density as seen by the hard parton for $\rho(\zeta)$, which is 
the local flow rapidity and $\alpha(\zeta)$, the  angle between the hydrodynamical
flow and the parton propagation direction. For a shower parton $a$, created at a 
time $\tau^0_a$ and evolving during $\tau_a$ before branching into a pair of 
offspring partons, the fully-integrated virtuality as propagated inside 
the shower code can be obtained from (\ref{eq:qhat}) and 
is given by
\begin{equation}\label{eq:deltaq2}
\Delta Q^2=\int_{\tau^0_a}^{\tau^0_a+\tau_a}d\zeta\hat{q}(\zeta), 
\end{equation}
which widens the available phase space from $Q^2\to Q^2+\Delta Q^2$ and therefore, 
the probability for extra (medium-induced) radiation. The integration in Eq.~(\ref{eq:deltaq2}) 
is taken over the eikonal trajectory of the parton-initiated shower 
from the production vertex to the exit from the medium. As explained in the 
introduction and also in~\cite{Perez-Ramos:2014mna}, where more details are given 
on the {\sc YaJEM} code description, the QCD splitting functions in the BW prescription are enhanced 
in the infrared sector by the medium parameter $f_{\rm med}$ which is related to the hydrodynamical
evolution sketched above~\cite{Borghini:2005em}. The dimensionful parameter
$K$ in Eq.~(\ref{eq:qhat}) characterizes the strength of 
the coupling between partons and the medium which can be obtained by tuning the 
measured hadron suppression factor $R_{AA}(p_t)$ with the RHIC data
in central 200 GeV AuAu collisions (see  Ref.~\cite{Renk:2009nz}).

Although this analysis is similar to the one used in~\cite{Perez-Ramos:2014mna} 
for the computation of fragmentation functions and its ratios, we explain 
the main steps in this section. The initial $p_T^{\rm parton}$ distribution 
of gluon and quark jets produced in pp (PbPb) collisions are 
simulated by sampling the convolution product of the (nuclear) parton distribution 
functions (n)PDFs with the matrix elements of the partonic hard scattering 
cross-section at 2.76 TeV. The nPDFs and PDFs are provided by the EKS
\cite{Eskola:1998df} and CTEQ \cite{Lai:2010nw} global-fits for heavy ions and hadron-hadron
collisions in the medium and vacuum respectively. The starting random selection of 
200000 dijets with center-of-mass energy $\sqrt{s}\sim2p_T^{\rm parton}$ on the
intervals $100\leq p_T^{\rm parton}({\rm GeV})\leq 120$, $120\leq p_T^{\rm parton}({\rm GeV})\leq 150$, 
$150\leq p_T^{\rm parton}({\rm GeV})\leq 300$ and $100\leq p_T^{\rm parton}({\rm GeV})\leq 300$ as input to 
{\sc YaJEM} and {\sc pythia6} is convenient for obtaining as small  
uncertainties and more accurate predictions as possible in the final 
comparison of the ratios with the CMS data. The next 
step involves the reconstruction of jets by using the anti-$k_t$ algorithm 
\cite{Cacciari:2011ma,Cacciari:2005hq,Cacciari:2008gp} for each $p_T^{\rm parton}$ range inside the 
jet cone of resolution $R=0.3$ with charged 
particles only, as in the CMS experiment. 

Reconstructed jets can be sorted by 
$p_T^{\rm jet}$ ($p_{T1}^{\rm jet}>p_{T2}^{\rm jet}>\ldots$) for the analysis such that the most hardest one 
($p_{T1}^{\rm jet}$) can be randomly selected from its ``almost" back-to-back pair ($p_{T2}^{\rm jet}$) 
event-by-event. The final cuts applied to each $p_T^{\rm parton}$ range, so as to 
match the experiment trigger selection, are 
$p_T^{\rm jet}\geq100$ GeV, $p_T^{\rm jet}\geq120$ GeV, $p_T^{\rm jet}\geq150$ GeV 
and $p_T^{\rm jet}\geq100$ GeV respectively. The theoretical fractions of gluon 
jets ``$f_{\rm g}$" following directly from the initial distribution of partons were given
in~\cite{Perez-Ramos:2014mna} for each $p_T^{\rm parton}$ range and found to be 
$\sim30\%$ in average. However, after the trigger selection is applied,
these fractions in each sample decrease dramatically, and 
especially for much narrower $p_T^{\rm parton}$ ranges such as $100\leq p_T^{\rm parton}({\rm GeV})\leq 120$
and $120\leq p_T^{\rm parton}({\rm GeV})\leq 150$ where the resulting
fraction due to the jet selection $p_T^{\rm jet}\geq100$ GeV is biased to $10^{-4}$~\cite{Perez-Ramos:2014mna}.
For each sample, we construct the mixed inclusive transverse momentum ($p_t$) 
distribution of hadrons and the ratio given by,
\begin{equation}
\label{eq:mixeddist}
\left(\frac{dN^{\rm h}}{dp_t}\right)_{\rm mixed}=f_{\rm g}\frac{dN^{\rm h}_{\rm g}}{dp_t}+
(1-f_{\rm g})\frac{dN^{\rm h}_{\rm q}}{dp_t},\qquad r=\left(\frac{dN^{\rm h}}{dp_t}\right)
^{\rm med}_{\rm mixed}\Big/\left(\frac{dN^{\rm h}}{dp_t}\right)^{\rm vac}_{\rm mixed}-1.
\end{equation}
such that hadroproduction is enhanced for $r>0$ and suppressed for $r<0$. Experimentally,
the ratios have been displayed as $({\rm PbPb}/{\rm pp})-1$ by the CMS 
collaboration~\cite{Chatrchyan:2014ava}.
\begin{figure}[!htbp]
\begin{center}
\epsfig{file=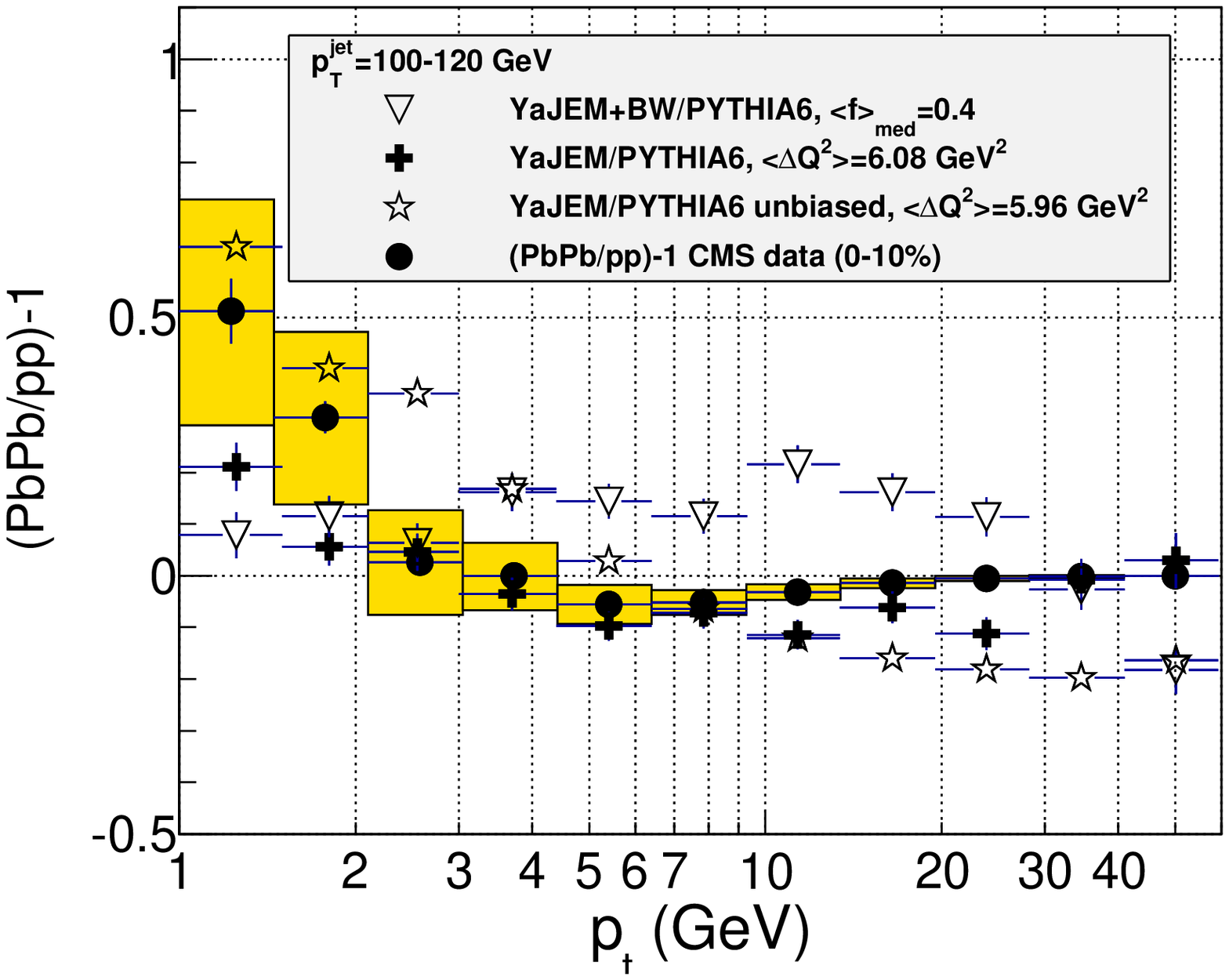, height=8.0truecm,width=8.4truecm}
\epsfig{file=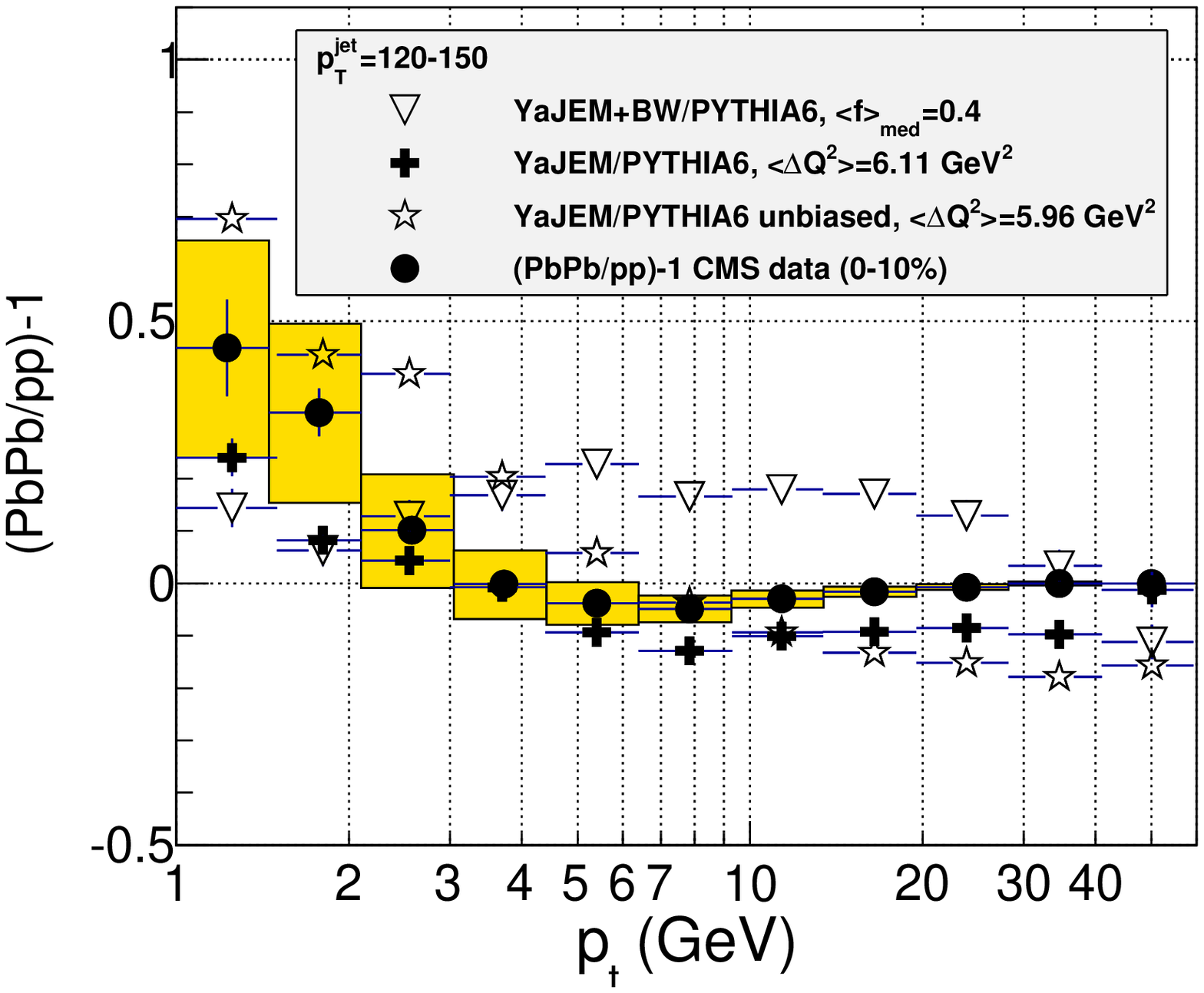, height=8.0truecm,width=8.4truecm}
\caption{\label{fig:pT1and2}  Comparison of hadron $p_T^{\rm jet}$ distribution ratios in PbPb over pp collisions
for jets with $100\leq p_T^{\rm jet} ({\rm GeV})\leq120$ (left) and 
$120\leq p_T^{\rm jet} ({\rm GeV})\leq150$ measured by CMS~\cite{Chatrchyan:2014ava} 
and obtained in two MC approaches ({\sc YaJEM} and {\sc YaJEM+BW}) as a 
function of the hadron's $p_t$.}  
\epsfig{file=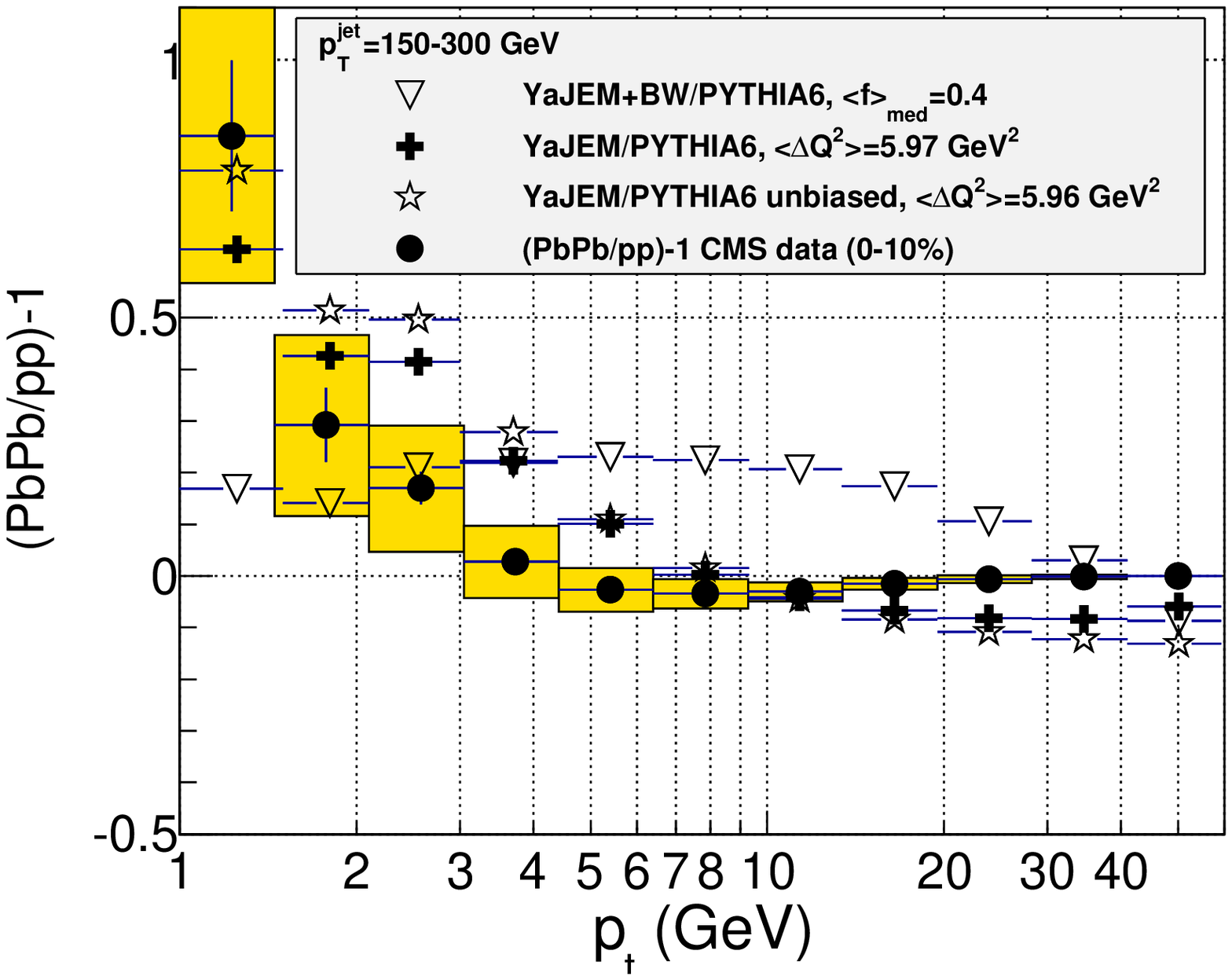, height=8.0truecm,width=8.4truecm}
\epsfig{file=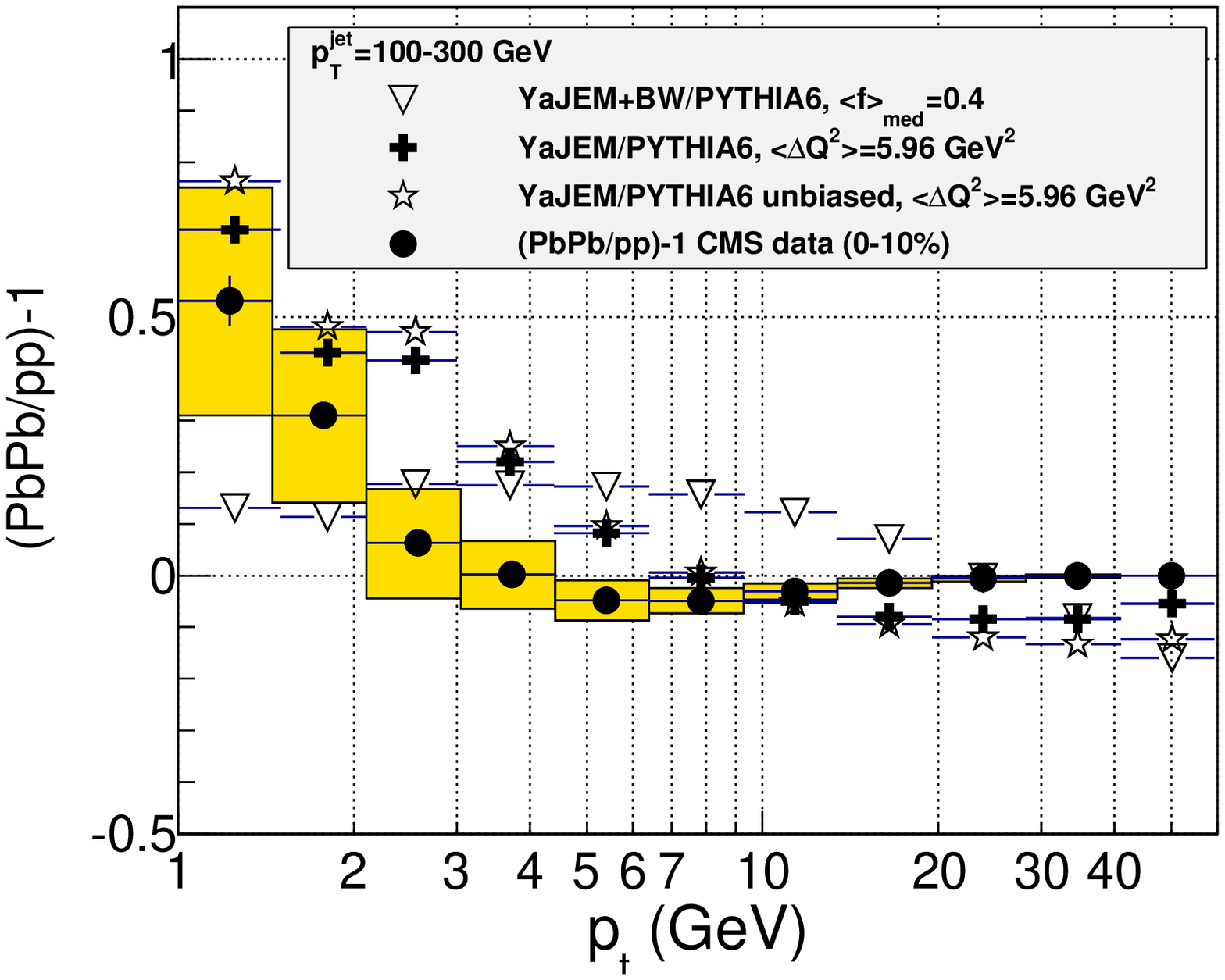, height=8.0truecm,width=8.4truecm}
\caption{\label{fig:pT3and4} Comparison of hadron $p_T^{\rm jet}$ distribution ratios in PbPb over pp collisions
for jets with $150\leq p_T^{\rm jet} ({\rm GeV})\leq300$ (left) and 
$100\leq p_T^{\rm jet} ({\rm GeV})\leq300$ measured by CMS~\cite{Chatrchyan:2014ava} 
and obtained in two MC approaches ({\sc YaJEM} and {\sc YaJEM+BW}) as a 
function of the hadron's $p_t$.}  
\end{center}
\end{figure}
In Figs.~\ref{fig:pT1and2} and \ref{fig:pT3and4}, we compare the final
results of our MC simulations, for the mixed ratios $r$ obtained from {\sc YaJEM/pythia6} for
$\langle\Delta Q^2\rangle\sim6$ ${\rm GeV}^2$ and {\sc YaJEM+BW/pythia6}
for $\langle f_{\rm med}\rangle\sim0.4$ in the BW model, with the CMS data 
in each $p_T^{\rm jet}$ range after averaging over a large amount of events in the 
hydrodynamical medium. As for the fragmentation functions discussed
in~\cite{Perez-Ramos:2014mna}, the {\sc YaJEM+BW/pythia6} ratios fail at
describing the shape and thereby, the physical features of 
the jet quenching phenomena in this framework. The biased and unbiased
{\sc YaJEM/pythia6} are displayed in the same panels together with the 
data. The unbiased ratios are softer at small $p_t$ (i.e. below 10 GeV) 
for all $p_T^{\rm jet}$ ranges and much harder than the experimental ratios especially for 
$100\leq p_T^{\rm jet}({\rm GeV})\leq 120$ and $120\leq p_T^{\rm jet}({\rm GeV})\leq 150$. 
In Fig.~\ref{fig:pT3and4}, the biased ratios {\sc YaJEM/pythia6} are closer to the CMS data, but the 
difference between the biased and unbiased ratios is not as large as in 
the other cases. 

\section{Summary}
In this paper we compared the inclusive hadron $p_t$ distributions computed with the {\sc YaJEM} and 
{\sc pythia6} MC simulations with recent CMS PbPb and pp jet data in the $p_T^{\rm jet}$ ranges 
$100\leq p_T^{\rm jet}({\rm GeV})\leq 120$, $120\leq p_T^{\rm jet}({\rm GeV})\leq 150$, 
$150\leq p_T^{\rm jet}({\rm GeV})\leq 300$ and $100\leq p_T^{\rm jet}({\rm GeV})\leq 300$ measured
at 2.76 TeV. The physical scenario implemented 
in {\sc YaJEM} describes qualitatively the data and reaches a much better agreement than the alternative 
{\sc YaJEM+BW} approach in $100\leq p_T^{\rm jet}({\rm GeV})\leq 120$, $120\leq p_T^{\rm jet}({\rm GeV})\leq 150$ despite
uncertainties caused by the account of different biases on the jet fragmentation analysis. 
For $150\leq p_T({\rm GeV})\leq 300$ and $100\leq p_T({\rm GeV})\leq 300$,
the ratios present an offset which makes the absolute medium-modified $p_t$ 
spectrum of hadrons softer at small $p_t$ and harder at large $p_t$ but
even more pronounced for unbiased showers. As for fragmentation functions~\cite{Perez-Ramos:2014mna}, 
the analyses provides a mean medium transport coefficient $\hat{q}\sim2.4$ 
${\rm GeV}^{2}/{\rm fm}$ with the obtained $\Delta Q^2\sim6$ ${\rm GeV}^{2}$ 
for a medium of length $L=2.5$ fm in this hydrodynamical description of the QGP.
Nevertheless, the comparison with other parton-energy-loss event generators
such as {\sc jewel}~\cite{Zapp:2008gi}, {\sc pyquen}~\cite{Lokhtin:2005px} 
and {\sc q-pythia}~\cite{Armesto:2009fj} should further constrain the medium
parameters and shed more light on the intra-jet transverse momentum 
structure and its interaction with the QCD medium formed in heavy-ion collisions.

\section*{Acknowledgments}

I strongly thank David d'Enterria for useful 
discussions and comments on the manuscript. I am also very thankful
to M. Cacciari, G.P. Salam and the LPTHE in Paris for their support during the preparation of these
results. 

\bibliographystyle{h-elsevier3}
\bibliography{mybib}

\end{document}